\documentclass[useAMS,usenatbib]{mn2e}

\usepackage{graphicx}
\usepackage{aas_macros}
\usepackage{times}

\title[Dust Formation in Mira Variables]
{Observable Effects of          Dust Formation in 
    Dynamic Atmospheres of M-type
Mira Variables}

\author[M.J.~Ireland and M.~Scholz]{M.J. Ireland$^1$\thanks{Present Address: Division of
    Geological and Planetary Sciences, Mail Code 150-21, California
    Institute of Technology, 1200 East California Boulevard, Pasadena,
    CA 91125, USA; mireland@gps.caltech.edu}, M. Scholz$^{1,2}$\\
$^1$School of Physics, University of Sydney NSW 2006, Australia\\
$^2$Institut f\"{u}r Theoretische Astrophysik der Universit\"{a}t Heidelberg,
Albert-Ueberle-Str.2, 69120 Heidelberg, Germany}

\begin{document}


\pagerange{\pageref{firstpage}--\pageref{lastpage}} \pubyear{2005}

\maketitle

\label{firstpage}

\begin{abstract}
 The formation of dust with temperature-dependent
 non-grey opacity is considered in a series of self-consistent model
 atmospheres at different phases of an O-rich Mira
 variable of mass 1.2~$M_\odot$. Photometric and interferometric
 properties of these models are predicted under different physical
 assumptions regarding the dust formation. The iron content of the initial
 silicate that forms and the availability of grain nuclei are found
 to be critical parameters that affect the observable
 properties. In particular, parameters were found where dust would
 form at 2-3 times the average continuum photospheric radius. This
 work provides a consistent physical explanation for the larger
 apparent size of Mira variables at wavelengths shorter than
 1\,$\mu$m than that predicted by fundamental-mode pulsation models.
\end{abstract}

\begin{keywords}
techniques: interferometric -- stars: variables: Miras -- 
stars: AGB and post-AGB
\end{keywords}

\section{Introduction}

  Dust shells around Mira variables often dominate the mid- and 
  far-infrared regions of their spectra and strongly influence the
  observed brightness distributions of their stellar and circumstellar
  material. In contrast to C-type Miras for which
  substantial theoretical progress has been made in recent years
  (e.g. \citealt{Helling00,Schirrmacher03,Hofner03,GautschyLoidl04}), 
  for oxygen-rich M-type Miras the physical conditions and the
  geometric region of formation of dust  
  grains is poorly understood.

  This situation is first of all due to the much more complex physics of
  formation of typical dust grains, which may contain metal oxides
  such as corundum (Al$_2$O$_3$), silicates or even solid iron
  \citep{Gail03}. These heterogenous grains form around nuclei that
  are in general composed of different compounds again, such as 
  TiO$_2$ \citep{Jeong03}.  
  Furthermore, in order to understand the occurrence of dust in an 
  M-type Mira, the formation of grains has to be studied within the 
  physical environment of the time-dependent dynamic atmosphere of the 
  pulsating star. Interferometric observations indicate that the inner 
  radii of dust shells may be as small as 2 to 4 stellar radii where 
  the stellar radius refers to the position of the continuum-forming 
  layers
  (e.g. \citealt{Danchi94,Danchi95,Lopez97,Monnier04,Ireland04a,Ireland05}).
  This is is well within the upper regions of the stellar atmosphere in 
  which typical molecular features of the stellar spectrum are
  formed (cf. \citealt{Hofmann98,Tej03b,Tej03,Ohnaka04a,Perrin04b,Ohnaka05})

  So far existing dynamic model atmospheres of M-type Mira stars
  either are dust-free (e.g. 
  \citealt{Hofmann98,Woitke99,Hofner03}),
  or they use grey, composition independent dust opacities
  (e.g. \citealt{Winters00,Jeong03}). 
  \citet{Bedding01} studied schematically the conditions of dust
  formations in the models of \citet{Hofmann98} and 
  concluded that corundum and silicate grains are likely to occur, 
  depending on the phase of pulsation, in the upper layers of the
  star's atmosphere. However, only a small number of 
  observable features of a small number of models were considered,
  with a very simple opacity treatment.  

  In this paper we investigate the formation of dust in a series of
  dynamic models of a Mira 
  variable, and discuss which observable effects would result from
  different physical assumptions regarding the dust formation.
  In particular, we focus
  on effects of non-grey opacities of silicate dust with a variable Fe
  content and the effect of grain size on observable properties.
  As hetrogenous O-rich dust formation is complex, our 
  approximations are designed to be most valid for only the innermost
  dust that has a significant influence on observable properties. We
  aim to parametrize uncertainties in this first significant dust
  formation with the smallest number of physical free parameters.

\section{Simple Dust Models}
\label{sectDustMod}

In this section, we will describe the procedure we used to
calculate absorption and scattering coefficients for silicate dust. We
will then discuss the degree of validity of this procedure for the
physics applicable to Mira atmospheres, the way in which
corundum dust is added to the models and finally the definition of the
dust types used in the models. 

The properties of a given dust type are
considered to be wholly determined by the dust temperature $T_d$ and
the gas pressure $P$. $T_d$ is found by solving the equation of
radiative equilibrium:  

\begin{equation}
 \int \kappa_\nu J_\nu d\nu = \int \kappa_\nu B_\nu d\nu
 \label{eqnJB}
\end{equation}

where $J_\nu$ is the mean intensity, $B_\nu$ = $B_\nu(T_d)$ is the Planck 
function, and $\kappa_\nu$ = $\kappa_{\nu}(T_d,P)$ is the dust absorption
coefficient in the layer with local dust temperature $T_d$ and local gas 
pressure $P$. The mean intensity is found iteratively in the modelling
procedure 
described in Section~\ref{sectModPred}. The gas temperature $T_g$ does not
enter in to this equation because the dust energy exchange with the gas
is negligible compared to the dust energy exchange with the radiation
field (e.g. see Appendix~D of \citealt{Gauger90}).

\subsection{Silicate Dust Properties}

We assume that the silicate that condenses first is olivine,
Mg$_{2x}$Fe$_{2-2x}$SiO$_4$, and use a simplified version of the
non-equilibrium treatment of \citet{Gail99} to determine the
condensation fraction of Si, $f_{\rm Si}$, and the composition parameter $x$. 
$f_{\rm Si}$ is assumed to follow the following simple relationship:

\begin{equation}
  f_{\rm Si} = \left\{ \begin{array}{ll}
    0                                         & T_d > 1240 + 46\log{P}\\
    0.25                                      & T_d < 1215 + 46\log{P}\\
    \frac{1240 + 46 \log{P} - T_d}{100}       & {\rm Otherwise}
    \end{array}\right.
 \label{eqnDustfDefn}
\end{equation}

The stability limit of silicate in this equation comes from a
linear fit to the stability limit for forsterite shown in
\citet{Gail03} for solar metallicity. The upper limit for Si condensation of
0.25 was chosen somewhat arbitrarily, so that there was not
unrealistic grain growth in outer atmospheric layers, where low gas pressures mean
that the silicate condensation fraction is far from equilibrium. 
The dependence of the olivine stability temperature on $x$ is neglected in
  Equation~\ref{eqnDustfDefn}.  However, its inclusion would not cause any
  significant changes to this paper, as the lowest value of $x$
  considered is 0.775, where the difference in stability temperature
  of the olivine from pure forsterite is only 30\,K. More
sophistication was not warranted in this 
relationship because the true equilibrium condensation fraction is a
function of both $T_d$ and $T_g$, and the system is not in general
close to chemical equilibrium.

In practice, the Si condensation fraction $f_{\rm Si}$ is determined by the functions
$x(f_{\rm Si})$ and $T_d(x)$, given that the opacity of silicate around
1\,$\mu$m is a very strong function of the composition parameter
$x$. In equilibrium condensation calculations, $x \approx 1$ for
the initial condensate as it 
is energetically more favorable in the
bulk solid to have Mg rather then Fe anions. However, the pulsating
atmosphere of a Mira variable is certainly not one where chemical
equilibrium considerations apply. To find $x(f_{\rm Si})$, we consider $x$ to be
defined by the conditions during rapid grain growth, 
approximate $K_p=0$ in Equations (55-57) of \citet{Gail99} (a reasonable
approximation as seen by the near-constant $x(T_d)$ functions in
Figure~3 of that paper) and assume that the grain has the same
composition $x$ throughout. This gives:

\begin{eqnarray}
  x &=& \frac{p_{\rm Mg}}{p_{\rm Fe} + p_{\rm Mg}} + \frac{\alpha}{2}
  \sqrt{\frac{m_{\rm SiO}}{m_{\rm Mg}}}\frac{p_{\rm Mg}}{p_{\rm SiO}} \\
    &=& \frac{\epsilon_{\rm Mg} - 2xf_{\rm Si}\epsilon_{\rm Si}}
       {\epsilon_{\rm Mg} + \epsilon_{\rm Fe}- 2f_{\rm Si}\epsilon_{\rm Si}} +  
      \frac{\alpha}{2}\sqrt{\frac{m_{\rm SiO}}{m_{\rm Mg}}} 
       \frac{\epsilon_{\rm Mg} - 2xf_{\rm Si}\epsilon_{\rm Si}}
       {\epsilon_{\rm Si}(1-f_{\rm Si})}
  \label{eqnDustxDefn}
\end{eqnarray}

Here $p_{\rm Mg}$ and $p_{\rm Fe}$ are the partial pressures of Mg and
Fe left in the gas mixture,
and $m_{\rm SiO}$ and $m_{\rm Mg}$ are the masses of an SiO molecule
and an Mg atom respectively. Equation~\ref{eqnDustxDefn} is solved for
$x$, which is limited to be less than or equal to 1. The elemental
abundances $\epsilon$ come from \citet{Palme81}, for
consistency with the modelling code. $\alpha$ is the
ratio of the exchange coefficient for 
Mg and Fe ions to the sticking coefficient for SiO molecules.

For $\alpha$ greater than or equal to 1, the assumption that the
grain has the same composition $x$ throughout is justified if the
diffusion timescale for Fe and Mg anions is small compared to the
timescale for grain growth. From the discussion in Section~5.2 of
\citet{Gail03}, this is the case for olivine at
dust temperatures above about 900\,K and grain sizes less than about
0.1$\mu$m. However, for small values of $\alpha$, the value of $x$ calculated
using Equation~\ref{eqnDustxDefn} will only be the composition of
newly deposited material on a growing grain, and not the composition
of the whole grain. During grain destruction,
Equation~\ref{eqnDustxDefn} does not accurately describe the grain
composition, but the general property predicted by this equation that
$x$ will increase as the grain evaporates is consistent with
experimental studies \citep{Ozawa00}.

In practice, Equation~\ref{eqnDustxDefn} is a good approximation
for $x$ during initial condensation or a phase of rapid grain growth
if $\alpha$ is greater than about 1.0. In other cases, it represents a
simple parameterization of $x$ that includes an appropriate decrease
in $x$ as Mg is depleted from the gas phase.

In forming Equation~\ref{eqnDustxDefn} from Equation~57 of
\citet{Gail99}, it was assumed that $\alpha_{\rm Mg} v_{\rm Mg} = \alpha_{\rm Fe}
v_{\rm Fe}$, the same assumption used to create Figure~3 of
\citet{Gail99}. In this relationship, the $\alpha_{\rm Mg}$ and $\alpha_{\rm Fe}$ 
parameters are sticking coefficients and the $v$ parameters are thermal
velocities. This is a necessary assumption given the
lack of experimental data for olivine sticking coefficients with
intermediate values of $x$.

Absorption and scattering coefficients are scaled as
in the Rayleigh limit, where at a constant condensation fraction the
absorption coefficient is independent of grain radius, and the
scattering coefficient is proportional to the cube of the grain
radius, $a$. The grain radius relates to the Si condensation fraction
$f_{\rm Si}$ and the number of available nuclei per hydrogen atom $N_{\rm nuc}$
by the formula:

\begin{equation}
 a = (\frac{3 \epsilon_{\rm Si} V_0}{4 \pi f_{\rm Si} N_{\rm nuc}})^{1/3}
 \label{eqnGrainRad}
\end{equation}

Here $\epsilon_{\rm Si}$ is the abundance of Si at solar metallicity, and
$V_0$ is the volume of the monomer, taken to be $7.5 \times 10^{-23}$
cm$^3$. The optical constants of the olivine with $x=0.5$ are taken from
\citet{Dorschner95}, and those for forsterite with $x=1.0$ taken from
\citet{Jager03}. These constants are used to derive $\kappa_{0.5}$, $\sigma_{0.5}$,
$\kappa_{1.0}$ and $\sigma_{1.0}$ in the Rayleigh limit for 30\,nm grains, with the
subscripts 0.5 and 1.0 representing $x$ for the olivine. The absorption
and scattering coefficients for the olivine are then approximated by a
simple interpolation:

\begin{eqnarray}
  \kappa_{\rm ol} &=& f_{\rm Si}((2-2x)\kappa_{0.5}
  +(2x-1)\kappa_{1.0}) \label{eqnKappa}\\
  \sigma_{\rm ol} &=& \frac{2.6 \times 10^{-11}}{N_{\rm nuc}} 
   f_{\rm Si}^2((2-2x)\sigma_{0.5}  + (2x-1)\sigma_{1.0})
\label{eqnSigma}
\end{eqnarray}






\subsection{Dust Formation in a Typical Mira Atmosphere}
\label{sectMiraDustFormation}

In the above parameterisation of absorption and scattering coefficients,
four important aspects of grain growth and destruction were not
discussed: dust nucleation, the approximation of using a single grain
radius $a$, the timescale of grain growth and
destruction compared to the pulsation timescale and the possibilities
for further grain growth beyond $f_{\rm Si}=0.25$. We will discuss these
one at a time, with specific reference to the physics of a Mira atmosphere.

The first stage of dust formation is the nucleation of seeds in a
super-saturated gas. The number of seed nuclei available for grain
growth determines the grain size. In the detailed nucleation discussion of
\citet{Jeong03}, the seed nuclei are found to form from
TiO$_2$. The nucleation rate is a complicated function of temperature
and pressure: at a typical Mira atmosphere dust formation pressure of
0.1\,dyn/cm$^{-2}$, the TiO$_2$ nucleation rate increases from $10^{-29}$ to
$10^{-17}$ per H atom per second as temperature decreases from 1200 to
1150\,K. This relationship has significant uncertainties, as it is
based on ab initio calculations from \citet{Jeong00} which predicts
an erroneous zero-point energy of TiO$_2$ by 0.73\,eV or 5\%, much larger
than experimental errors (compare Table~2 of that paper with
\citet{Balducci85}). If one assumes all energies are erroneous by
  this relative amount, then errors in nucleation temperatures of the
  order of 50\,K would be expected. Given these difficulties, we
will consider that grain nucleation is complete when the growth of
dominant dust components begins, and leave the number of grain nuclei
per H atom $N_{\rm nuc}$ a free parameter in our models.

By separating nucleation and growth in this artificial way, we make
the approximation that all dust grains in a single atmospheric layer
are the same size. The effect of this dust on absorption and scattering is only
equivalent to a distribution of particle sizes in the Rayleigh limit when
the grain radius $a << \lambda$, a condition always met for the dust
types and wavelengths considered here. This approximation is also more
desirable than fixing a single particle size distribution at all
atmospheric layers, as that approach doesn't include the increase in
scattering coefficients with increasing grain radius.

As material cooling behind a shock in a Mira atmosphere moves
  outwards, the dust condensation fraction given by
  Equation~\ref{eqnDustfDefn} rapidly increases as the central
  luminosity decreases near minimum and the radiation field becomes
  increasingly geometrically diluted. For the dust treatment here
  (assuming $f_{\rm Si}$ is a function of the instantaneous variables $T_d$ and
  $P$), the grain growth must be rapid compared to the dynamical
  timescale. According to \citet{Gail99}, the limiting factor in the growth rate of
olivine is the deposition of an SiO molecule on the grain
  surface. Considering a super-saturated gas mixture at 1100\,K with
  an assumed sticking
coefficient for SiO of 0.1, the rate of grain growth is greater than
  10\,nm in 0.1 cycles ($3 \times 10^6$\,s) whenever $\log(P) >
  -2.22$. Therefore, we shall be cautious in interpreting any dynamical  
grain growth and destruction effects that occur at $\log(P) < -2.22$.

At dust temperatures where olivine has mostly condensed, there are a
variety of possibilities for further dust condensation, including
quartz, and solid iron. In a non-equilibrium scenario, further
model condensation will depend on the initial choice of $\alpha$, as well as
other unknown parameters. Note that at the near-maximum model phases,
$f_{\rm Si}=0.25$ is never reached in the outer layers (near the arbitrary
5~$R_p$ model surface: see Section~\ref{sectModPred}), and the low
gas pressures would mean that any growth would be slow compared to the
dynamical timescale.

\subsection{Addition of Corundum}

It has been suggested (e.g. \citealt{Maldoni05}; \citealt{Egan01})
that corundum (Al$_2$O$_3$) plays a significant role in radiative
transfer processes 
in O-rich AGB stars. This is due to the clear need for opacity in the
12-15\,$\mu$m range when fitting models to observed spectral energy
distributions. However, physical models of Mira atmospheres with
realistic gas densities (e.g. \citealt{Bedding01}) have previously
shown that the opacity of corundum and the abundance of Al are not
high enough to have a significant effect on observed
spectra. Therefore, we will re-examine the effects of corundum in
this paper.

The difficulty in modelling the growth of corundum grains is that
nucleation and growth will certainly be simultaneous, due to the high
corundum stability temperature. In this paper we are not concerned
about the details of this process, but attempt to allow corundum to
have the maximum plausible effect on the radiative transfer.

We have taken the stability limit for corundum at solar metallicity
  from \citet{Gail03}, fitting a function linear in $\log{P}$ for
 $-4 < \log{P} < 0$. For corundum dust only, the Al condensation
  fraction $f_{Al}$ is given by: 

\begin{equation}
 f_{Al} = \left\{ \begin{array}{ll}
    0                                         & T_d > 1600 + 60\log{P}\\
    1.0                                       & T_d < 1500 + 60\log{P}\\
    \frac{1600 + 60 \log{P} - T_d}{100}       & {\rm Otherwise}
    \end{array}\right.
 \label{eqnCorundumDustfDefn}
\end{equation}

The dust temperature $T_d$ is found using Equation~1, with the optical
constants of corundum taken from \citet{Koike95} and the Rayleigh 
limit again assumed. For silicate dust, corundum absorption and
scattering coefficients are also added to those of the silicates, with
$f_{Al}=f_{\rm Si}$. This procedure is designed to approximate the
effect of heterogenous grains with a corundum core and a silicate
mantle. Any pure corundum grains would have a much lower $T_d$ in the
outer atmosphere than a grain that includes silicate with $x < 1$, and
therefore would not be significant if hetrogenous grains are also
present.

Possibly, the optical constants of \citet{Koike95}
  over-estimate the absorption in the near-infrared region of the
  spectrum, given that their optical constants $k$ for amorphous
  corundum are high when compared to thin-film alumina \citep{Harris55} or
  crystalline corundum \citep{Harman94}. However, a lower value for
  $k$ could not cause a $> 20$\% change in the calculated stability
  radius of corundum, as the gas is already becoming optically-thick
  in the mid-IR at the calculated stability radius due to molecular
  absorption, meaning that the dust can not have a significantly lower
  temperature than the gas. Furthermore, a smaller stability radius
  for corundum would place it at radii smaller than those at which
  successful nucleation could occur
  (cf. Section~\ref{sectMiraDustFormation}).


\subsection{Model Dust Types}

The dust types considered in this paper are given in
Table~\ref{tblDustTypes}. Dust types A and B have $\alpha=1.0$, and
represent olivine that will form pure forsterite initially in an
  outward-moving gas packet, before
becoming enriched by Fe as the partial pressure of Mg
decreases at $f \approx 0.15$. A higher value of $\alpha$ would give
dust where Mg condenses more fully before the addition of Fe to
the olivine. The difference between dust A and B is the number of
grain nuclei, and hence the grain size: dust A will have dust grains
$\sqrt[3]{10}$ times larger than dust B. Dust C is a dust where some
the initial condensate contains an appreciable amount of Fe ($x=0.93$), giving a
higher absorption coefficient at wavelengths from 1 to 4\,$\mu$m,
where the bulk of the stellar flux 
is emitted. 

Dust type D, with corundum only, can be
thought of as the dust that would exist if $\alpha < 0.4$, as in this
case silicates do not condense in our models. The effect of corundum
on radiative transfer processes is at
roughly the maximum plausible with the chosen value of $N_{\rm nuc} =
10^{-13}$, because much larger grains could not grow in an
outward-moving gas packet even with sticking coefficients of
1.0.


\begin{table}
 \caption{Dust types considered in this paper. Columns are: Code used
 in this paper; Types of dust included (Olivine and Corundum); log of
 the number of grain nuclei per H nucleus; and $\alpha$, the ratio of
 the exchange coefficient for Mg and Fe to the sticking coefficient
 for SiO (see text). }
 \begin{tabular}{@{}llll@{}}
  \hline
    Dust Code & Dust Types & $\log(N_{\rm nuc})$ & $\alpha$ \\
  \hline
    A         & Ol         & -13.3               & 1.0   \\
    B         & Ol         & -12.3               & 1.0   \\
    C         & Ol         & -12.3               & 0.63  \\
    D         & Cor        & -13                 & -     \\
  \hline
 \end{tabular}
 \label{tblDustTypes}
\end{table}



\section{Modelling Details and Results}
\label{sectModPred}

  The atmospheric models used in this study are based on model 
  series of \citet{Hofmann98}, supplemented by models of
  \citet{Tej03} and \citet{Ireland04d,Ireland04c}. 
  These are non-grey dynamic model atmospheres for an M-type Mira
  variable with fundamental-mode period 332 days, mass 1.2 $M_\odot$ 
  (M series; P series: 1.0 $M_\odot$) and solar metallicity.
  The luminosity $L_p$ and Rosseland radius $R_p$ of the non-pulsating 
  parent star are 3470 $L_\odot$ and 260 $R_\odot$ (M series;
  P series: 241 $R_\odot$). We refer to \citet{Hofmann98} for 
  essential details of model construction.  In particular, these are
  self-excited models that were allowed to relax over a very large
  number of cycles. Pulsation is not strictly
periodic, and the successive cycles that were studied in detail by 
Hofmann et al. (1998) and Ireland et al. (2004b, c) were so chosen that
cycles of the M and P series with different characteristics are included. Phases 
assigned to these models (Ireland 2004b, c) are based on a reasonably chosen
mean zero point, and the individual phase of a model of a specific cycle
always refers to this zero point. The hydrodynamic models were calculated with 
dust-free grey opacities, and a non-grey temperature stratification of the 
atmospheric layers was then obtained in a second step by solving the 
radiative-equilibrium equation (i.e. equation 1 with $B_\nu(T_g)$, gas 
absorption coefficient $\kappa_{{\nu}g}(T_g,P)$, gas temperature $T_g$) with
dust-free non-grey opacities (see Hofmann et al. 1998), retaining the 
gas-pressure stratification of the hydrodynamic model.

  During pulsation, the position of the
  continuum-forming layers oscillates around $R_p$ with an amplitude
  of $\pm 30$\%. The parameters for the models at
  the individual phases considered here are shown in
  Table~\ref{tblMSer}, reproduced from \citet{Ireland04d}. For
  comparisons with observations at specific visual phases, 0.12 will
  be added to $\phi_{vis}$ in Table~\ref{tblMSer} \citep{Ireland04d}.
  
 Using these dust-free models as a starting point, the dust types as
  described in Section~\ref{sectDustMod} were added to the models, and the
  temperature profile, spectra and intensity profiles
  re-calculated. The new gas-temperature $T_g$ and the
  dust-temperature $T_d$ stratifications were obtained by iteration,
  that is, radiative equilibrium was enforced for both components (gas
  and dust) in each iteration step for obtaining for the next step new
  values of gas and dust temperatures, of gas and dust absorption
  coefficients, and of mean intensities $J_\nu$ (cf. comments on
 dust type D2 in \citealt{Bedding01}). Note that scattering
  is approximated as being isotropic and coherent in these models and,
  hence, does not appear explicitly in Equation~1. Note also, that
  radiative acceleration does not 
become sufficiently large in the dusty models for generating a dust-driven 
wind, so that retaining the gas-pressure stratification of the original grey
dynamical model is a reasonable approximation.

\begin{table}
 \caption{Parameters of M series Mira models. The
  columns: visual phase $\phi_{vis}$; luminosity $L$; 1.04~$\mu$m
  near-continuum radius $R_{1.04}$; and the effective
  temperature $T_{1.04}$ corresponding to 
  $R_{1.04}$.}
\begin{tabular}{@{}lllll@{}}
 \hline
  Model  & $\phi_{vis}^1$ & $L$            & $R_{1.04}$ & $T_{1.04}$ \\
        &                & ($L_{\odot}$)  & ($R_p$)    & (K)        \\ 
 \hline 
M05  & 0+0.49& 1470 & 0.84 & 2420 \\ 
M06n & 0+0.60& 2430 & 0.78 & 2860 \\
M08  & 0+0.77& 4780 & 0.81 & 3320 \\
M09n & 0+0.89& 5060 & 1.03 & 2970 \\
M10  & 1+0.02& 4910 & 1.18 & 2760 \\
M11n & 1+0.11& 4360 & 1.21 & 2640 \\
M12n & 1+0.21& 3470 & 1.18 & 2540 \\
M12  & 1+0.27& 2990 & 1.12 & 2500 \\
M14n & 1+0.40& 1670 & 0.91 & 2400 \\
M15  & 1+0.48& 1720 & 0.83 & 2530 \\
M16n & 1+0.60& 2460 & 0.77 & 2860 \\
M18  & 1+0.75& 4840 & 0.81 & 3310 \\
M18n & 1+0.84& 4980 & 1.00 & 3010 \\
M19n & 1+0.90& 5070 & 1.09 & 2900 \\
M20  & 2+0.05& 4550 & 1.20 & 2680 \\
M21n & 2+0.10& 4120 & 1.21 & 2610 \\
M22  & 2+0.25& 2850 & 1.10 & 2490 \\
M23n & 2+0.30& 2350 & 1.03 & 2460 \\
M24n & 2+0.40& 1540 & 0.87 & 2410 \\
M25n & 2+0.50& 2250 & 0.79 & 2780 \\
 \hline
\end{tabular}
\newline $^1$ 0.12 should be added to these phases to give a more accurate
model phase. See \citet{Ireland04d}.
 \label{tblMSer}
\end{table}

 The M series was chosen over the P series for this study because
  approximations at 3-5\,$R_p$ caused unrealistic
  density jumps with respect to time in the P series.  
 These approximations included coarse gridding in the dynamical 
 models, and an artificial density-gradient cutoff that had been enforced in the
 outer layers at some phases/cycles for computational reasons (non-physical 
 density discontinuity at the 5\,$R_p$ model surface). 
 The M series models considered here have no artificial
  cutoff applied, and have a fine enough gridding of the dynamical
  models at 3-5\,$R_p$ to produce no unrealistic effects.

  The P and M model series are meant to describe typical Miras like
  $o$~Cet or R Leo. Comparison with observations of both stars
  \citep{Hofmann01,Ireland04b,Ireland04d,Woodruff04,Fedele05}
  show satisfactory agreement
  of observed and predicted features, but various deviations are also
  obvious from the discussion of \citet{Ireland04c}.
  We may check in this study whether some of these deviations are
  due to the omission of dust in the original models. Note that the
  mean-opacity treatment of strong water bands cools the outer layers
  too much and back-warms the continuum photosphere, as evident in the
  spectra of \citet{Tej03}. If the forest of water lines were treated
  correctly in the models,  
  one would expect the modelled water-bands to become less deep and
  the continuum (J-K) colour to become redder. This would give colour
  temperatures slightly lower  
  than $o$~Cet or R~Leo. For this reason, we expect that realistic dust 
  formation in these models would be at least as efficient as 
  in $o$~Cet or R~Leo.
    


\begin{figure}
 \includegraphics{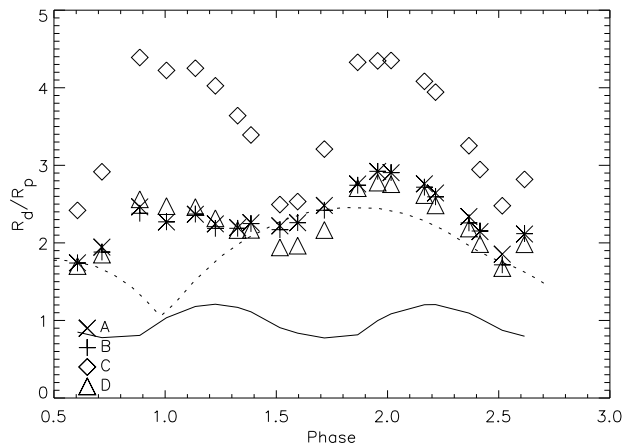}
 \caption{Model dust formation radii plotted against phase (0.12 has been added to
 the Table~\ref{tblMSer} phases - see text) for the
 four dust types considered here. The solid line shows the position
 where the dust-free near-continuum radius $R_{1.04}$, and the dotted line 
 shows the position of a single mass zone as a function of phase.}
 \label{figDustRadii}
\end{figure}

Figure~\ref{figDustRadii} shows the dust stability radii for all dust
types in Table~\ref{tblDustTypes}. The dotted line shows the location
of a single mass zone (the upper mass zone in Figure~2 of \citet{Hofmann98}),
giving an indication of the dynamics of the 
atmosphere. For dust types A, B and D, by far the largest rate of
grain growth occurs just after initial grain formation, so it is in
the initial stages of grain growth where it is most important to
examine our assumption of dust condensation responding instantly to
changes in the radiation field. For the mass zone shown in
Figure~\ref{figDustRadii} and zones immediately above it, the largest
rate of grain growth occurs between phases 1.3 and 1.7, where the
material crosses the dust stability radius. At these
phases the material behind a strong shock (which reaches $2.75 R_p$
at phase 1.7) is cooling as it moves outwards. 
As the material crosses the dust stability radius for dust types
A and B at these phases, $\log(P)$
is between $-1.9$ and $-1.7$ with a local pressure scale 
height of $0.3 R_p$. At these pressures,
dust growth is sufficiently rapid according to the considerations of
Section~\ref{sectDustMod}. However, this condition is not well-satisfied
at all phases, so the details of the phase-dependence of the
dust condensation radii should not be considered a reliable prediction
of these simple models.

\begin{figure}
 \includegraphics{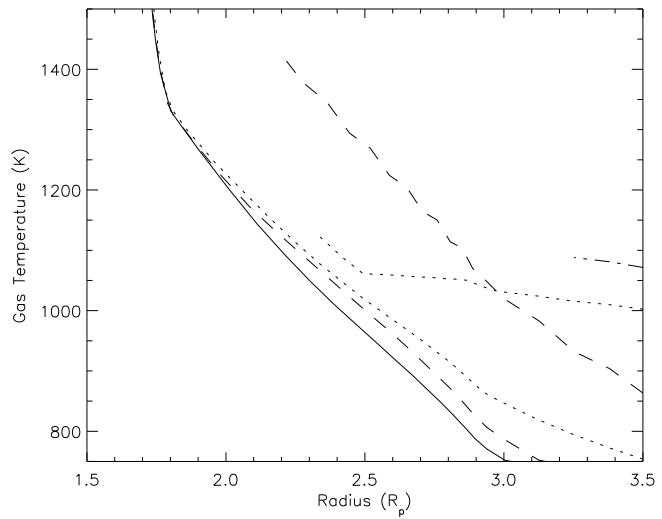}
 \caption{Gas temperatures (lower lines) and dust temperatures (upper
 truncated lines) for the M22 models. The solid line is for
 the dust-free model, the dotted line dust A, the dashed line dust
 D and the dot-dashed line dust C. Dust type B is nearly
 indistinguishable from the dotted line and the gas temperature
 for dust C is nearly indistinguishable from the solid line. The dust-free
 near-continuum radius $R_{1.04}$ for this model is 1.1~$R_p$.}
 \label{figGasTemp}
\end{figure}

\begin{figure}
 \includegraphics{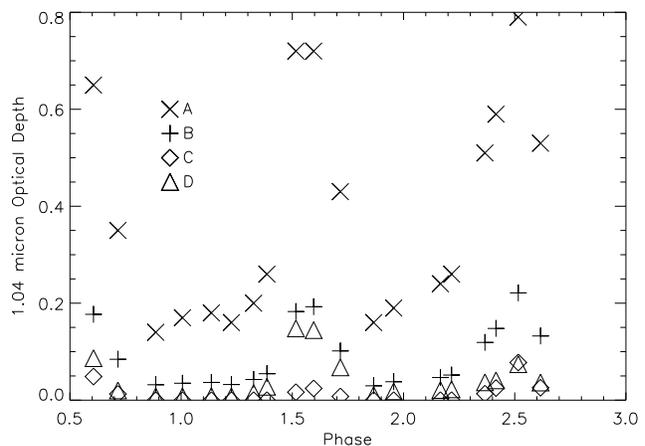}
 \caption{Total optical depths at 1.04\,$\,u$m as a function of phase
 for the four dust types.}
 \label{fig1umODepth}
\end{figure}

The total optical depth of the dust at 1.04\,$\mu$m as a function
of phase is shown in Figure~\ref{fig1umODepth} for all dust types. The large optical
depths for dust type A are due to scattering by grains that reach a
maximum radius of 96\,nm 
in the outer layers, and reach 80\,nm radius rapidly after
condensation while $x$ is equal to 1.0. The optical-depths for dust
types A and B imply that the $1.04$\,$\mu$m near-continuum wavelength
may not be a good window for interferometrically observing the
continuum-forming photosphere as suggested by
e.g. \citet{Jacob02}. As scattering opacity scales as
$\lambda^{-4}$ in the Rayleigh limit, the J and H band windows are not
nearly as affected by the presence of these dust types as the
1.04\,$\mu$m window. It is clear that dust types C
and D have would have only a minimal effect at this wavelength.

The C-rich dust models of \citet{Hofner03}, at nearly twice the
luminosity of the M-series considered here, have a large
radiative acceleration of the dust grains which significantly modifies
the dynamic stratification of these models. At the 5~$R_p$ surface of the
models considered here, the radiative acceleration ranges from 0.08 to
0.29 times gravitational acceleration for dust A. This range of values
is smaller for all 
other dust types or for radii less than 5~$R_p$.
This is easily
understandable as dust only forms within 5~$R_p$ if the dust has a low
opacity between 1 and 4 microns, where 
the bulk of the stellar flux is emitted. Therefore, we do not expect
radiative acceleration of dust grains to play a highly significant role in the dynamic
stratification of a typical Mira variable of luminosity less than
about $5000 L_{\odot}$ within 5 parent-star radii. However, beyond the
surface of the models presented here radiative acceleration is expected
to become significant as long as it is possible for the near-infrared
opacity of the dust to increase with further Fe condensation at the
low pressures ($< 10^{-4}$\,dyn/cm$^2$) encountered at these radii.


Figure~\ref{figGasTemp} shows the influence of these dust types on the
gas temperature for the M22 model. The gas in outer layers is warmed
by about 100\,K for dust types A and B.  Back-warming isn't
significant for any dust type in the layers where most spectral
features are formed, which is at temperatures above about
1400\,K. Other models showed effects similar to those in this
model. In this case, and in most other models, the dust
  temperatures are higher than the gas temperatures. This provides
  further justification for assuming that nucleation is complete when
  grain growth begins. The sharp change in slope of the dotted line at
  2.5~$R_p$ is due to the the presence of zero Fe ($x=1.0$) for smaller
  radii, and an increasing fraction of Fe at larger radii.


\section{Observable Dust Effects}

The presence of all dust types considered here has a minimal effect on
photometry between a wavelength of 1 and 9\,$\mu$m. Near minimum, the
J band flux is reduced  by 0.1 mag due to extinction by dust for dust
types A and B, with no measurable effects on H and K band 
photometry. However, this small J-band effect is not a reliable model
prediction, as it requires growth and destruction rates in the outer
layers near minimum that are too high for outer layer pressures
$P<10^{-3}$\,dyn/cm$^2$. Up to 0.3 mag
extinction for dust type A occurs near-minimum at 900\,nm
due to back-scattered radiation re-intercepting the lower atmosphere. The
extinction of V band light by the dust considered here is too
difficult to model, both due to the difficulties in treating strong
TiO absorption, and to the influence of the grain size
distribution on extinction at short wavelengths.

\begin{table*}
 \caption{Model predictions for fitted Gaussian FWHMs to the
 stellar intensity profile in milli-arcsec
 for various wavelengths, for the M22 model which is at a phase of
 0.37.  Measurements for
 R~Leo and $o$~Cet for 4 separate instruments at visual phases between 0.2 and
 0.49 are also shown (see text).}
 \begin{tabular}{@{}llllllllll@{}}
  \hline
    Wavelength (nm) & No Dust & Dust A & Dust B & Dust C & Dust D & WHT & AAT & SAO & COAST\\
  \hline
    700     &  22.1  & 50.1 & 27.3 & 22.2 & 23.3 &  27.6/38.2 & 38  & 32.7 & -  \\
    750     &  21.8  & 46.9 & 25.6 & 21.8 & 22.7 &  -         & 32  & 30.5 & -  \\
    920     &  22.3  & 37.3 & 24.1 & 22.4 & 22.7 &  23.0      & 23  & -    & 29.3 \\
    1045    &  17.8  & 26.7 & 18.7 & 17.8 & 18.1 &  -         & -   & 23.6 & - \\
  \hline
 \end{tabular}
 \label{tblVisDiams}
\end{table*}

Interferometric observations at wavelengths short-wards of about 1\,$\mu$m
are sensitive to scattering from dust. Without allowing for
  significant effects of dust or molecules, larger than
expected apparent diameters have led some authors
(e.g. \citealt{Haniff95}) to favour overtone pulsation for Mira
variables. This proposal is clearly in conflict with the now conventional
 understanding \citep{Wood99,Scholz00} that Miras pulsate in the
 fundamental mode.
Note that this discrepancy also existed in the near-infrared
\citep{vanBelle96}, but correctly including the effects of molecular
contamination of continuum diameters in physical models has since
provided a resolution in this case \citep{Woodruff04,Fedele05}. 
Several authors have clearly mentioned the possibility of 
dust scattering increasing the apparent size of Mira variables
(e.g. \citealt{Danchi94,Hofmann01,Ireland04a}), but a physical
model is required to discriminate correctly between the effects of TiO
absorption and dust.

Table~\ref{tblVisDiams} shows
measured diameters of $o$~Cet and R~Leo from several experiments
compared with model predictions of the M22 model with different dust
types. The observation phases and targets were: $o$~Cet at phases 0.47
and 0.49 (different cycles) for the William Herschel Telescope (WHT)
data \citep{Tuthill95}, $o$~Cet at phase 0.34 for the Anglo-Australian
Telescope (AAT) data \citep{Ireland04a}, 
R~Leo at phase 0.2 for the Special Astrophysical Observatory (SAO)
data \citep{Hofmann01}, and R~Leo at phase 0.32 from Cambridge Optical Aperture
Synthesis Telescope (COAST) \citep{Burns98}. The COAST data was presented as
uniform-disk fits, and the correction factor of 1.52 in the text of
\citet{Burns98} was applied. 

For the model fits in Table~\ref{tblVisDiams}, a least-squared
Gaussian fit to visibility 
$V$ (the normalised Fourier amplitude of the source brightness
distribution) was calculated for baselines shorter than where
$V=0.3$. A Gaussian fit was both cited as a better fit than a uniform
disk by the authors of the observational papers, and is a better fit
in general for the dusty models described here. The model star was
placed at 105\,pc, in-between the K-band 
maximum fit distances for R~Leo and $o$~Cet for the M series
\cite{Ireland04d} and consistent with {\em Hipparcos} distances. 
Note that by having models that match the observed temperature as
  measured by continuum J-K colours \citep{Ireland04d} and using the
  K-band fit distance, the comparison 
  between observed and predicted angular diameters is not heavily
  dependent on this distance. For example, if the models are
  under-luminous by 20\% with radii 10\% too small, then the model
  stars will be placed at a distance 10\% too close and the predicted
  angular diameters will remain unchanged.

For the 700, 750 and 920\,nm filters, a
20\,nm bandwidth was assumed, and for 1045\,nm a monochromatic
prediction is given. Although these bandwidths did not match 
the individual experimental bandwidths exactly, the experimental filters all
included the chosen wavelengths and would be expected to have similar
fit diameters according to the measurements of \citet{Ireland04a}. 

It is clear that amongst these models, the observed large diameters
and the increase in observed diameter at shorter wavelengths can
only be produced by dust A or B, with an intermediate dust being
favorable. The scatter in the observations is indicative of
  both the slightly different phases of observations, and
  cycle-to-cycle variations in the atmospheric structure. This is also
seen in the models, where the fitted FWHMs for the M12 model would be
smaller than those for the M22 model.

The wavelengths of 700, 750, 920 and 1045\,nm were chosen for this
comparison because they are not sensitive to strong TiO absorption in the
upper atmosphere. In order to exclude TiO absorption from causing
the larger apparent diameters, we have carefully examined the
treatment of TiO opacity in our models.
The M series spectra have TiO absorption band depths
that are much too deep when compared to observations. Although one
significant reason for this in the figures of \citet{Tej03} was
out-dated chemical data for TiO$_2$ (since replaced with data from
\citealt{Sharp90}), a remaining large problem in modelling the deep
absorption bands 
correctly are non-LTE effects in TiO band formation in a dynamic
atmosphere. From the TiO line list of \citet{Schwenke98}, the typical
Einstein~A coefficient for the upper level in a strong TiO absorption
band is $10^7$\,s$^{-1}$. This is much higher than the collision rate
in the atmospheric regions
where the TiO features are formed ($P < 1$\,dyn/cm$^2$ and $T <
2000$\,K). Several simple attempts have been made by us to
characterize the magnitude of non-LTE effects. Indeed, spectra with
band depths near that observed for typical Miras can be produced by
assuming an ad hoc TiO temperature profile. Although these tests
affected the strong TiO features significantly, the regions of the
spectrum such as those in Table~\ref{tblVisDiams} were largely
un-affected. This is because these regions of weak and
strongly temperature-dependent absorption are formed significantly
deeper in the atmosphere than the strong absorption bands.
Therefore, we are quite confident that the large apparent
 diameters can not be caused by TiO alone, and that therefore dust
opacity is required to produce the observed diameters.

The Optical Interferometric Polarimetry (OIP) observations of
\citet{Ireland05} are generally consistent with the predictions
here. The radii of the dust shells in this paper are approximately 
2.5~$R_p$, when the model star is placed at the distance corresponding
to that which fits the K band maximum. Fitted optical-depths to these
observations are between those predicted from the type A and type B
dusts. However, the two stars observed in that paper (R~Car and
RR~Sco) have smaller K-band photometric amplitudes than the
M-series, and have much less dust emission than $o$~Cet or
R~Leo. Therefore, one might expect the gas densities for these Miras
to be smaller than that in the M series at the same radii. This would
mean that grain 
formation and destruction may not be fast compared to the pulsation at
the grain formation radii, and would need a more detailed,
time-dependent study.


\begin{figure}
 \includegraphics{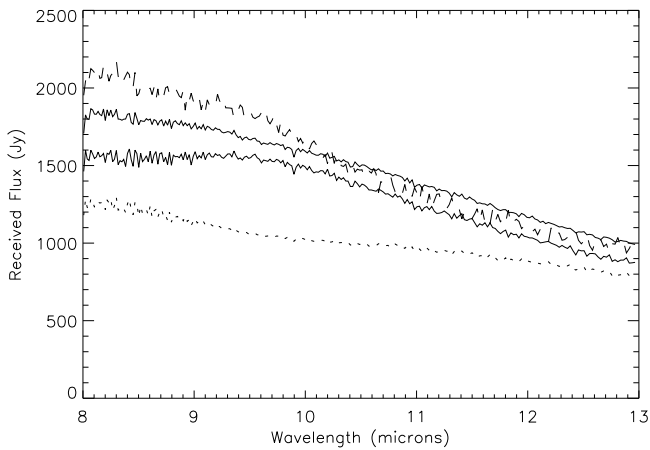}
 \caption{Model spectra for B dust between 8 and 13 microns for the M11n model
 (lower solid line), M15 model (dotted line), M18 model (dashed line)
 and M21n model (upper solid line).}
 \label{fig8_13Spect}
\end{figure}

In order to most accurately model the mid-infrared
spectra, a new spectrum and intensity profile computation code was
constructed. This code used as inputs the gas pressure and molecule
partial pressures, the gas and dust temperature and the velocity
stratification from the model-construction code.
It takes into account continuum opacities, H$_2$O lines from
\citet{Partridge97}, TiO lines from 
\citet{Schwenke98} and other diatomic lines from the input to the ATLAS12 models
\citep{Kurucz94}. The contributions to line profiles from
different layers were redshifted or blueshifted as appropriate
according to the difference in projected velocities. A
micro-turbulence of 2.8\,km/s was assumed, in between than used in the 
M-giant models of \citet{Plez92} and that derived by \citet{Hinkle79}.

The mid-infrared spectra are significantly influenced by dust types A and
B, but the spectra for dust types C and D are similar to
dust-free models. This is due to the low condensation fraction of dust
C at radii where densities are high, and for dust D this is due to the
low abundance of
Al when compared to Si, Fe and Mg. As dust types A
and B result in similar condensation 
fractions but different grain sizes, their effects on the mid-infrared
spectra are very similar. Hence, only selected mid-infrared spectra
for representative models with dust B are shown in
Figure~\ref{fig8_13Spect}, smoothed to have a resolving power of
500. The model star is placed at a distance of 102\,pc, to match the
fit distance for the $o$~Cet K-band light curve \citep{Ireland04d}. 

The most crucial point to note about these spectra is that they underestimate
the measured flux from $o$~Cet at all wavelengths by a factor of about
2 to 3, based on the range of photometry in \citet{Monnier98}. This is
consistent with the measurements of \citet{Danchi94} and 
\citet{Weiner03}, whose data require approximately
half to three quarters of the total 
mid-infrared flux from $o$~Cet to originate from a region further from
the central star than our model surface (which corresponds to a 58\,mas
radius at a distance of 102\,pc). This dust is almost certainly part
of an outflow from this star, and may have a shell-like structure as
seen in outflows from other AGB stars (e.g. \citealt{Hale97}).
Indeed, in the models a strong shock front occasionally (no more than
about 1 cycle in 4) drives a significant amount of mass through
the 5~$R_p$ surface where it may form part of a stellar wind. 
Therefore, this additional flux is not inconsistent with the models
  here, it is simply not part of the models due to the arbitrary
  5~$R_p$ model surface.

\begin{figure}
 \includegraphics{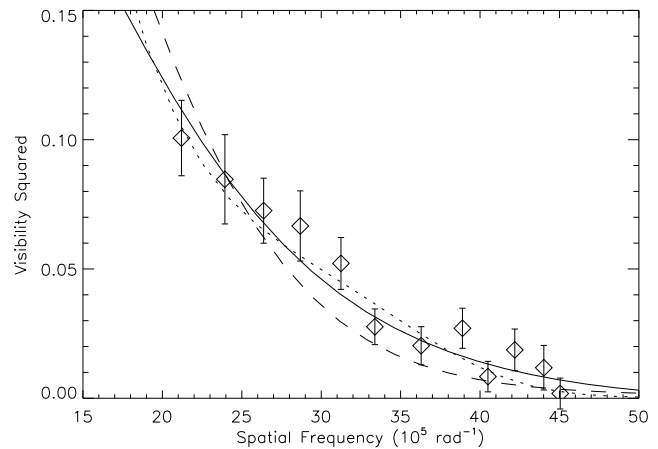}
 \caption{Visibility data used in \citet{Weiner03} at 
  a pulsation phase of 0.20 (data from November 28,
  2000). One-parameter fits (see text)
  for models at similar 
  phases (see text) are shown for the M11n model (solid line) and the
  M21n model (dashed line). The dotted line illustrates a
  three-parameter fit (see text) based on the 
  M21n model and an additional shell of previously ejected dust.}
 \label{WeinerComp}
\end{figure}

A second important point to note about these model spectra is that
they are not well approximated by a black-body spectrum and a silicate
emission peak. According to the classification scheme of
\citet{Sloan98}, the spectra for the M series with dust B could be
classified from SE2 to SE8, with a dust emission contrast of 0.14 to
0.3. This suggests that the observed mid-infrared excess for some Mira
variables may not be strictly associated with an outflow
(e.g. \citealt{Ireland05}). 



The effect of these dust types on mid-infrared interferometry is
difficult to predict. This is due to the large fraction of the light
that originates from dust further than the 5~$R_p$ surface
to these models \citep{Weiner03}. Nevertheless, one can say that a
general feature of the models is that they do not resemble a
uniform-disk intensity profile. Figure~\ref{WeinerComp} shows
11.15\,$\mu$m $V^2$ data from $o$~Cet observations at a phase of 0.2
obtained from J. Weiner (personal communication). This is calibrated data which was
used for the model fits in \citet{Weiner03}, unlike the un-calibrated
data plotted in Figure~1 of that 
paper. Intensity profiles were computed using dust B, and 
the narrow observing bandwidth, assuming a redshift of
83\,km/s. Over-plotted in Figure~\ref{WeinerComp} are the model
predictions of the M11n 
model, at a pulsation phase of 0.23, and the predictions of the M21n
model at a pulsation phase of 0.22. The model star was placed at the
fit distance for the $o$~Cet K-band light curve
\citep{Ireland04d} and the fraction of emission from an over-resolved
component of extended dust emission was a free parameter. This
  fraction has the effect of multiplying the squared visibility by a
  constant smaller than 1.

Certainly in the case of the M11n model the fit is
reasonable, and the 45\,\% additional flux required from extended dust
emission is roughly consistent with observed mid-infrared excesses for
$o$~Cet. In the case of the M21 model, additional model parameters
would be required to fit the data reasonably.
The example plotted as the dotted line is one where two additional parameters were
  free: the radius of an additional shell of dust and the fractional
  emission from this shell. 
 The parameters for this model were a dust
  shell radius of 5.83~$R_p$ containing 18\,\% of the total 
emission. This demonstrates that for mid-infrared interferometry to
provide accurate constraints on the innermost dust and water
emission, a model or observations of dust at large radii must be
combined with models of the inner regions as described here. Examining
data from several other epochs did not give significant errors when fitting
dust B models to the data, but dust-free models were not
in general extended enough to fit the data from \citet{Weiner03}.

A final observable feature of the dust formation models as described
here is the phase- and cycle-dependence of the dust formation. A
relatively quick increase in dust opacity is predicted near minimum
during certain cycles wherever a shock front passes through the radius
of dust formation. This occurs around phase 1.4 in these models, where
the 1.04\,$\mu$m dust optical depth rapidly increases from 0.26 to
0.72 for dust A. This increase in opacity would be best seen
interferometrically by a jump in size at either
wavelengths short-wards of 1\,$\mu$m or in the mid-infrared. Combined 
observations at these two wavelength regimes could constrain grain sizes
(e.g. the difference between dust types A and B).

The cycle-dependence of the outer layers in the M and P series relates
to their chaotic nature. There is no reason to
expect these layers to show high degrees of spherical symmetry as
assumed in the models. A crude approximation is that one
could think of opposite sides of a star to be in different
model `cycles'. Similar behaviour is evident on large scales in
  the models of \citet{Woitke05}. 
For this reason, dust formation near-minimum may also be 
observable as the creation of asymmetric features in a shell 
at 2-3~$R_p$. One effect of this dust creation would be rapid changes in
the angle of polarization observed. This effect is a plausible
explanation for the rapid changes in polarization angle observed
around Julian day 2441000 (near phase 0.5) for R~Leo by
\citet{Serkowski01}. These
observations show polarizations of several percent
in U, B and V bands where the angle of polarization changes by 25
degrees in one direction over 50 days, then 50
degrees in the opposite direction over the following 50 days.

\section{Conclusion}

Condensation of corundum and silicate dust in a
model of a 1.2~$M_\odot$ self-excited Mira variable has been
examined. Four dust types were modelled, representing approximated
physical dust formation parameters.  
It was found that a dust type between A and B from
Table~\ref{tblDustTypes} fits existing observations well
\footnote{Predictions of existing and additional models are
  available upon request to anyone interested for specific
  observational programs.}. In
particular, the effect of scattering by this type of dust explains the
systematically large observed apparent diameters at wavelengths shorter than
1\,$\mu$m.  This dust type has approximately $10^{-13}$ grain nuclei per H 
atom available at the onset of silicate condensation, and requires
$\alpha$, the ratio of the exchange coefficient for Mg and Fe ion
to the sticking coefficient of SiO, to be at least the order of unity. A model where
only corundum forms was found to be a poor fit to observations of
$o$~Cet and R~Leo. The use of composition-dependent opacities enabled
the survival of the initial Mg-rich silicate condensate at radii of
2-3 times the model parent star radius.

\section*{Acknowledgments}
This research was supported by the Australian Research Council and
the Deutsche Forschungsgemeinschaft within the linkage project ``Red
Giants'', and by a grant of the Deutsche Forschungsgemeinschaft
on "Time Dependence of Mira Atmospheres". We would like to thank
P.~G. Tuthill and the referee P.~Woitke for valuable comments on the
manuscript, and H.~P. Gail for answering many questions about dust
formation theory. 
 
\bibliography{../mireland}
\bibliographystyle{../mnras/mn2e}

\label{lastpage}

\end{document}